\documentclass[aps,prb,superscriptaddress,showpacs,floatfix]{revtex4-1}
\usepackage{graphicx,amsmath,amssymb,xspace,epsfig,float,multirow,subfigure,tabularx}

\pdfoutput=1

\begin{document}

\title{Superconductivity in 2D electron gas induced by high energy optical
phonon mode and large polarization of the STO substrate.}
\author{Baruch Rosenstein}
\email{vortexbar@yahoo.com}
\affiliation{Electrophysics Department, National Chiao Tung University, Hsinchu 30050,
\textit{Taiwan, R. O. C}}
\affiliation{Physics Department, Ariel University, Ariel 40700, Israel}
\author{B.Ya. Shapiro}
\email{shapib@biu.ac.il}
\affiliation{Physics Department, Bar-Ilan University, 52900 Ramat-Gan, Israel}
\author{I. Shapiro}
\affiliation{Physics Department, Bar-Ilan University, 52900 Ramat-Gan, Israel}
\author{Dingping Li}
\email{lidp@pku.edu.cn}
\affiliation{School of Physics, Peking University, Beijing 100871, \textit{China}}
\affiliation{Collaborative Innovation Center of Quantum Matter, Beijing, China}

\begin{abstract}
Pairing in one atomic layer thick two dimensional electron gas by a single
flat band of high energy longitudinal optical phonons is considered. The
polar dielectric $SrTiO_{3}$ (STO) exhibits such an energetic phonon mode
and the 2DEG is created both when one unit cell $FeSe$ layer is grown on its
$\left( 100\right) $ surface and on the interface with another dielectric
like $LaAlO_{3}$ (LAO). We obtain a quantitative description of both systems
solving the gap equation for $T_{c}$ for arbitrary Fermi energy $\epsilon
_{F}$, electron-phonon coupling $\lambda $ and the phonon frequency $\Omega $%
, and direct (RPA) electron-electron repulsion strength $\alpha $. The focus
is on the intermediate region between the adiabatic, $\epsilon _{F}>>\Omega $%
, and the nonadiabatic, $\epsilon _{F}<<\Omega $, regimes. The high
temperature superconductivity in 1UC$FeSe$/STO is possible due to a
combination of three factors: high LO phonon frequency, large
electron-phonon coupling $\lambda \sim 0.5$ and huge dielectric constant of
the substrate suppression the Coulomb repulsion. It is shown that very low
density electron gas in the interfaces is still capable of generating
superconductivity of the order of $0.1K$ in LAO/STO.
\end{abstract}

\maketitle


\section{Introduction}

Single layer of iron selenide ($FeSe$) grown on a strong polar insulator $%
SrTiO_{3}\left( 001\right) $ (STO) exhibits superconductivity\cite%
{Wang12,Jia15,ultrafast16,Wang14,Lee14,swave} at surprisingly high
temperature $70-100K$. This is an order of magnitude larger than the parent
bulk material with the superconducting transition temperature\cite{Hsu08} $%
T_{c}$ of $8K$. This suggests that the dominant mechanism of creation of the
superconductivity in the $FeSe$ layer might be different from that of the
bulk $FeSe$ and is caused by influence of the STO substrate. To strengthen
this point of view the high-resolution angle-resolved photoemission
spectroscopy (ARPES) experiments\cite{Lee14} and the ultrafast dynamics\cite%
{ultrafast16} demonstrated the presence of high-energy phonons in STO. The
frequency of the oxygen longitudinal optical (LO) mode reaches $\Omega
\approx 100meV$. In addition it turns out that the phonons couple strongly
to the electrons in the $FeSe$ layer (the coupling constant was estimated to
be\cite{ultrafast16} $\lambda \sim 0.5$, much larger than in the parent
material, $\lambda =0.19$). The band is flat with only a small momentum
transfer to electrons. This identification is supported by the earlier ARPES
on STO surface states, which shows a phonon-induced hump at approximately $%
100meV$ away from the main band and through inelastic neutron scattering\cite%
{phononSTO}. The role of substrate in assisting superconductivity is not
limited to generation of phonons. The polar STO has a huge dielectric
constant (estimated to be above $\epsilon =1000$ on the surface) and hence
suppresses Coulomb repulsion inside the $FeSe$ layer.

The nature of electronic states within the $FeSe$ layer is by now quite
settled experimentally. The Fermi surface of the single unit cell (1UC)
consists of two electron-like pockets centred around the crystallographic
M-point (Brillouin zone corners) with a band bottom below the Fermi level%
\cite{Lee14} $\epsilon _{F}=60meV$. This means that electrons form a two
dimensional electron gas (2DEG) with small chemical potential. The novelty
of the superconducting system is that the occupied states are close to the
band edge, very far from the classic case. In both conventional (BCS) and
unconventional superconductors the chemical potential is the largest energy
scale in the problem (even in quasi 2D high $T_{c}$ cuprates the chemical
potential is order of magnitude higher). STM measurements in the
superconducting state demonstrates that there are no nodes\cite{swave} (no
sign change of the order parameter). It shows at $4K$ a fully peaked gap
(with double peaks at $10mev,15mev$ with minimum at $5eV$) which is
suppressed only by magnetic impurities, similar to a conventional 2D s-wave
superconductor. Absence of nesting indicates that there are no effects like
charge density waves.

An early theory\cite{Lee12} focused on the screening due to the STO
ferroelectric phonons on antiferromagnetic spin fluctuations mediated Cooper
pairing in parent material $FeSe$. It suggested that the phonons
significantly enhance the Cooper pairing and even change the pairing
symmetry. Naively the spin fluctuation interaction by itself should lead to
nodeless d-wave pairing. For the electron-phonon coupling $\lambda \sim 1$
the enhancement was large, although perhaps not enough to explain the
experiment. When the inter-pocket electron-phonon scattering is also strong,
opposite-sign pairing will give way to equal-sign pairing. Later \cite{Lee14}
it was suggested that the interfacial nature of the coupling assists
superconductivity in most channels, including those mediated by spin
fluctuations.

Another idea \cite{Hirshfeld15} is to use both the electron pockets at the
Fermi surface band and the "incipient" hole band below it also found in
ARPES, namely generalizing to the multiband model. The conclusion was that
"a weak bare phonon interaction can be used to create a large $T_{c}$, even
with a spin fluctuation interaction which may be weakened by the incipient
band." The difficulty is that the forward scattering nature of the essential
phonon processes then means that LO phonons cannot contribute to the
inter-band interaction. Gor'kov considered\cite{Gorkov16} polarization on
the surface, screening and the STO surface LO phonon pairing. His conclusion
is that the LO phonon mediated pairing alone cannot account for
superconductivity at such high $T_{c}$.

The small chemical potential is typical for the STO systems. Another related
superconducting (with much lower $T_{c}$) 2DEG system with even much smaller
chemical potential is the $LaAlO_{3}$(LAO)-STO interface observed earlier%
\cite{InterfaceSCexp}. The microscopic origin of the superconductivity in
the LAO/STO system is already quite clear\cite{Mannhart15}. It is the BCS -
like s-wave pairing attributed to the same LO phonon modes discussed above
in context of the 1UC$FeSe/$STO system. Spin fluctuations seem not to play
any role in the pairing leading to superconductivity. The phase diagram of
LAO/STO is qualitatively similar to the dome-shaped phase diagram of the
cuprate superconductors: in the underdoped region the critical temperature
increases with charge carrier depletion.

The theoretical effort to understand the LAO/STO system\cite{Klimin14}
resulted in realization that the Migdal-Eliashberg theory of
superconductivity, valid when the phonon frequencies are much smaller than
the electron Fermi energy, should be generalized. This is not the case for
polar crystals like STO with sufficiently high optical-phonon frequencies,
and consequently the dielectric function approach proposed long ago by
Kirzhnits\cite{Kirzhnits} and developed in ref\cite{Takada} proved to be
useful. It was shown that the plasma excitations are important at larger $%
\mu $ (reduce the electron-phonon coupling) and enable to explain the
non-monotonic behavior of $T_{c}$ as function of bias that changes chemical
potential.

In this paper we further develop a theory of superconductivity in 1UC$FeSe$%
/STO and LAO/STO based on phononic mechanism including effects of the
screened Coulomb repulsion. In the \textit{first stage} a simple model of
2DEG with pairing mediated by a dispersionless LO phonons is proposed with
Coulomb repulsion assumed to be completely screened by huge polarization of
STO ($\epsilon \sim 3000$ in 1UC$FeSe$/STO). In this case the gap equations
of the Frohlich model can be reduced (without approximations) to an integral
equation with one variable only and are solved numerically for arbitrary
Fermi energy $\epsilon _{F}$, phonon frequency $\Omega $ and electron-phonon
coupling $\lambda <1$. An expression for the adiabatic and nonadiabatic
limits are derived and results for $T_{c}$ compare well with experiments on
1UC$FeSe$/STO. Then, in the \textit{second stage} we include the RPA
screened Coulomb repulsion (for somewhat smaller values of dielectric
constants are estimated\cite{Klimin12} to be $\epsilon =186$ on the STO side
and $\epsilon =24$ on the LAO side) and solve a more complicated gap
equations numerically (without making use of the Kirzhnits Ansatz) for
various $\epsilon _{F}$ and Coulomb coupling constant. Both the adiabatic, $%
\epsilon _{F}>>\Omega $, (conventional BCS) and the nonadiabatic, $\epsilon
_{F}<<\Omega $, cases are considered and compared with the local model
studied earlier in the context of BEC physics\cite%
{BEC2Drev,Melik,Randeria2D,Chubukov}. The Coulomb repulsion results in
significant reduction or even suppression of superconductivity. A
phenomenological model for dependence of $\epsilon _{F}$ and $\lambda $ on
electric field for the LAO/STO is proposed.

The paper is organized as follows. The basic 2DEG phonon superconductivity
model is introduced in Section II. The general Gaussian approximation for
weak electron-phonon interactions and RPA screening is described in Section
III. The superstrong screening case (neglecting Coulomb repulsion
altogether) case is solved Section IV. The same calculation is performed
using the  Kirzhnits  approach in Section V. The general case including the RPA
screened Coulomb repulsion is investigated numerically in Section VI. The
phenomenology of 1UC$FeSe$/STO and LAO/STO and comparison with experiments
are discussed in Section VII followed by Discussion and summary. Appendices
A and B contain the derivation of Gorkov equations and the 2D RPA
neutralizing background contribution respectively.

\section{The LO phonon model of pairing in 2DEG}

As mentioned above various STO systems including 1UC$FeSe$/STO (medium to
low density) and interface LAO/STO the (very low density) electron gas
appears localized in a plane of width of one unit cell (in FeSe layer or on
the STO side respectively). The Hamiltonian of the system contains three
parts%
\begin{equation}
H=H_{e}+H_{ph}+H_{e-ph}\text{.}  \label{Hamiltoniandef}
\end{equation}

\subsection{Description of 2DEG}

We use a continuum parabolic 2DEG model one "flavours" (up and down spin
projections and two valleys in 1UC$FeSe/STO$) with effective mass close to
mass of electron\cite{Klimin14}). The 2DEG Hamiltonian in terms of creation
operators $\psi _{\sigma }^{\dagger }\left( r,t\right) $, $\sigma =\left\{
\uparrow ,1\right\} ,...\left\{ \uparrow ,N\right\} ,\left\{ \downarrow
,1\right\} ,...\left\{ \downarrow ,N\right\} $\ electrons thus is
\begin{equation}
H_{e}=\int_{r}\psi _{\sigma }^{\dagger }\left( -\frac{\hbar ^{2}\nabla ^{2}}{%
2m}-\mu \right) \psi _{\sigma }+\frac{1}{2}\ \int_{r.r^{\prime }}n\left(
r\right) v\left( r-r^{\prime }\right) n\left( r^{\prime }\right) \text{,}
\label{He}
\end{equation}%
where the charge density operator is
\begin{equation}
n\left( r\right) =\psi ^{\sigma \dagger }\left( r\right) \psi ^{\sigma
}\left( r\right) \text{,}  \label{n}
\end{equation}%
and $\mu $ is the chemical potential (Fermi energy). The electron-electron
interactions, not related to the crystalline lattice, are described by
potential $v\left( r\right) $. The electrostatics on the surface/interface
is quite intricate\cite{Klimin12}, and we approximate it by the Coulomb
repulsion:
\begin{equation}
v\left( r\right) =\frac{e^{2}}{\epsilon r}\text{,}  \label{Vstat}
\end{equation}%
where $\epsilon $ is and effective 2D dielectric constant of the system. As
mentioned in Introduction the effective dielectric constant is huge in STO
at low temperatures due to the ionic movements.

\subsection{Phonons and electron-phonon interactions}

Crystal vibrations in STO are highly energetic. The single phonon band\cite%
{phononSTO,Mannhart15} near $\Omega =100meV$ is most probably associated
with pairing attractive electron-electron force is the ferroelectric LO that
involves the relative displacement of the $Ti$\ and $O$\ atoms. The high
energy STO oxygen LO phonon band mode is separated from all the other phonon
bands by a substantial energy gap\cite{phononSTO}. The single branch of the
optical phonons described by the bosonic field\cite{Fetter} $\phi \left(
r\right) =\sum\nolimits_{k}\frac{1}{\sqrt{2}}\left( b_{k}^{\dagger
}e^{-ikr}+b_{k}e^{ikr}\right) $. The phonon part of the Hamiltonian
therefore is:

\begin{equation}
H_{ph}=\frac{1}{2}\int_{r,r^{\prime }}\phi \left( r\right) v_{ph}\left(
r-r^{\prime }\right) \phi \left( r^{\prime }\right) \text{,}
\label{Hamiltonian}
\end{equation}%
where the phonon energy density $v_{ph}\left( r-r^{\prime }\right) $, for
the nearly flat LO band is approximately local:

\begin{equation}
v_{ph}\left( r\right) =\hbar \Omega \delta \left( r\right) \text{.}
\label{id}
\end{equation}%
Experiments demonstrated a substantial electron--phonon coupling $g$. In
fact the collective mode energy is greater or comparable to the width of the
electron band. Importantly, the electron--phonon coupling allows only a
small momentum transfer to the electron.%
\begin{equation}
H_{e-ph}=g\int_{r}n\left( r\right) \phi \left( r\right) \text{.}
\label{Heph}
\end{equation}

Despite the simplifications, the model is far from being solvable and
standard approximations are applied in the following section to obtain the
critical temperature of the superconductor. Various "bare" parameters like
effective masses, $\Omega $, the electron-electron and electron-phonon
couplings are renormalized as the interaction effects\ are accounted for.

\section{The pairing equations}

\subsection{Matsubara Action}

We use the Matsubara time $\tau $ ($0<\tau <\hbar /T$) formalism\cite{Fetter}
with action corresponding to the Hamiltonian Eq.(\ref{Hamiltoniandef}%
),(setting $\hbar =1$), $\ A\left[ \psi ,\phi \right] =A_{e}\left[ \psi %
\right] +A_{ph}\left[ \phi \right] +A_{e-ph}\left[ \psi ,\phi \right] ,$with

\begin{eqnarray}
A_{e} &=&\int_{r,\tau }\psi _{\sigma }^{\ast }\left( r,\tau \right)
D^{-1}\psi _{\sigma }\left( r,\tau \right) +\frac{1}{2}\int_{r,r^{\prime
},\tau }n\left( r,\tau \right) v\left( r-r^{\prime }\right) n\left(
r^{\prime },\tau \right)  \label{Action} \\
A_{ph} &=&\frac{1}{2}\int_{r,r^{\prime },\tau }\phi \left( r,\tau \right)
d^{-1}\phi \left( r^{\prime },\tau \right) ;  \notag \\
&&A_{e-ph}=g\int_{r,\tau }n\left( r,t\right) \phi \left( r,t\right) \text{.}
\notag
\end{eqnarray}%
\ Here the electron Green's function is,
\begin{equation}
D^{-1}=\partial _{\tau }-\frac{\nabla ^{2}}{2m}-\mu \text{,}  \label{iD}
\end{equation}%
while that of the phonon field is
\begin{equation}
d^{-1}=\left( -\partial _{\tau }^{2}+\Omega ^{2}\right) \delta \left(
r-r^{\prime }\right) \text{.}  \label{id1}
\end{equation}

In Fourier space the action reads
\begin{eqnarray}
A_{e} &=&\sum_{p\omega }\psi _{p\omega }^{\sigma \ast }D_{p\omega }^{-1}\psi
_{p\omega }^{\sigma }+\frac{1}{2}\sum_{p\omega p_{1}p_{2}\omega _{1}\omega
_{2}}v_{p}\psi _{p_{1}\omega _{1}}^{\sigma \ast }\psi _{p_{1}-p,\omega
_{1}-\omega }^{\sigma }\psi _{p_{2}\omega _{2}}^{\rho \ast }\psi
_{p_{2}+p,\omega _{2}+\omega }^{\rho };  \label{Action2} \\
A_{ph} &=&\frac{1}{2}\sum_{k\omega }\phi _{k\omega }^{\ast }d_{\omega
}^{-1}\phi _{k\omega };A_{e-ph}=g\sum_{pp_{1}\omega \omega _{1}}\psi
_{p_{1}\omega _{1}}^{\sigma \ast }\psi _{p_{1}-p,\omega _{1}-\omega
}^{\sigma }\phi _{p\omega }  \notag
\end{eqnarray}%
with electronic,
\begin{equation}
D_{p,\omega }^{-1}=i\omega +\varepsilon _{p};\text{ }\varepsilon
_{p}=p^{2}/2m-\mu \text{,}  \label{prop_e}
\end{equation}%
and optical phonon
\begin{equation}
d_{\omega }^{-1}=\frac{\omega ^{2}+\Omega ^{2}}{\Omega ^{2}}\text{,}
\label{propph}
\end{equation}%
propagators respectively. The fermionic Matsubara frequencies are $\omega
_{n}=\pi T\left( 2n+1\right) $, while for bosons $\omega _{n}=2\pi Tn$ with $%
n$ being an integer. In 2D%
\begin{equation}
v_{p}=\frac{2\pi e^{2}}{\epsilon p}\text{.}  \label{2DCoulomb}
\end{equation}%
The action can be treated with the standard gaussian approximation.

\subsection{The pairing equations \ }

The electronic action is obtained by integration of the partition function
over the phonon field,

\begin{equation}
Z_{e}\left[ \psi \right] =\int_{\phi }e^{-A\left[ \psi ,\phi \right]
}=e^{-A_{e}^{eff}\left[ \psi \right] }\text{.}  \label{action}
\end{equation}%
The gaussian integral is,%
\begin{eqnarray}
A_{e}^{eff}\left[ \psi \right] &=&\sum_{\omega p}\psi _{p\omega }^{\sigma
\ast }D_{p\omega }^{-1}\psi _{p\omega }^{\sigma }+  \label{Ae} \\
&&+\frac{1}{2}\sum_{\omega \omega _{1}\omega _{2}pp_{1}p_{2}}V_{p\omega
}\psi _{p_{1}-p,\omega _{1}-\omega }^{\sigma \ast }\psi _{p_{1}\omega
_{1}}^{\sigma }\psi _{p_{2}\omega _{2}}^{\rho \ast }\psi _{p_{2}-p.\omega
_{2}-\omega }^{\rho }\text{,}  \notag
\end{eqnarray}%
where $V_{p\omega }=V_{p\omega }^{RPA}+V_{\omega }^{ph}$. The part of the
effective electron-electron attraction due to phonons is:%
\begin{equation}
V_{\omega }^{ph}=-g^{2}\frac{\Omega ^{2}}{\ \omega ^{2}+\Omega ^{2}}\text{.}
\label{Vph}
\end{equation}%
To take into account screening, we made the replacement $v_{p}\rightarrow
V_{p\omega }^{RPA}$ (the random phase approximation) in 2D,

\begin{equation}
V_{p\omega }^{RPA}=v_{p}\left( 1+\frac{Nmv_{p}}{\pi }\left( 1-x/\sqrt{x^{2}+1%
}\right) \right) ^{-1}\text{, \ }  \label{VPRA}
\end{equation}%
where $x=\left\vert \omega \right\vert /\left( v_{F}p\right) $ with $%
v_{F}^{2}=2\mu /m$.

Performing the standard gaussian approximation averaging, see appendix A,
one arrives at the Gor'kov equations for the normal ,$\left\langle \psi
_{k\omega }^{\uparrow I\dagger }\psi _{q\nu }^{\downarrow J}\right\rangle
=\delta _{\omega -\nu }\delta _{k-q}\delta ^{IJ}G_{k\omega }$ ($I,J=1,...,N$
are flavours), and the anomalous, $\left\langle \psi _{k\omega }^{\uparrow
I}\psi _{q\nu }^{\downarrow J}\right\rangle =\delta _{\omega +\nu }\delta
_{k+q}^{{}}\delta ^{IJ}F_{k\omega }$, Greens functions. The result is%
\begin{equation}
-\Delta _{k\omega }^{\ast }F_{k\omega }+D_{k\omega }^{\ast -1}G_{k\omega }=1,
\label{Gor1}
\end{equation}%
and
\begin{equation}
\Delta _{k\omega }G_{k\omega }=-D_{k\omega }^{-1}F_{k\omega }\text{,}
\label{Gor2}
\end{equation}%
where the gap function is defined by%
\begin{equation}
\Delta _{k\omega }=\sum\nolimits_{p_{1}\omega _{1}}V_{p_{1}-k,\omega
_{1}-\omega }F_{p_{1}\omega _{1}}\text{.}  \label{deltadef}
\end{equation}%
Near the critical point one can neglect higher orders in $\Delta $ in Eq.(%
\ref{Gor1}), resulting in $G=D^{\ast }$. Substituting this into Eq.(\ref%
{Gor2}), one gets:%
\begin{equation}
\sum\nolimits_{p\nu }\left\vert D_{p\nu }\right\vert ^{2}V_{\mathbf{p-k},\nu
-\omega }^{{}}\Delta _{p\nu }=-\Delta _{k\omega }\text{.}  \label{gapeq}
\end{equation}%
Using the explicit form of the propagator $D$,$\,\ $Eq.(\ref{prop_e}), the
equation takes a final form:%
\begin{equation}
\sum\nolimits_{\mathbf{p}m}\frac{2NT}{\omega _{m}^{2}+\varepsilon _{p}^{2}}%
V_{\mathbf{p-k},m-n}^{{}}\Delta _{\mathbf{p}m}=-\Delta _{\mathbf{k}n}\text{.}
\label{gap1}
\end{equation}

\subsection{Simplification of the integral equations for critical
temperature for the s-wave pairing.}

Transforming to polar coordinates and using rotation invariance, $\Delta _{%
\mathbf{p}\nu }=\Delta _{p\nu }$, $p=\left\vert \mathbf{p}\right\vert $, and
then changing the variables to $\varepsilon _{p}=p^{2}/2m-\mu $, the
electronic part of the kernel of Eq.(\ref{gap1}) is

\begin{equation}
\int_{\varepsilon _{2}=-\mu }^{\Lambda -\mu }\frac{mNT}{\pi }%
\sum\nolimits_{n_{2}}\frac{1}{\omega _{n_{2}}^{2}+\varepsilon _{2}^{2}}%
P_{\varepsilon _{1}\varepsilon _{2};n_{1}-n_{2}}\Delta _{\varepsilon
_{2}n_{2}}=-\Delta _{\varepsilon _{1}n_{1}}\text{.}  \label{inteq}
\end{equation}%
Here $\Lambda $ is an ultraviolet cutoff of the order of atomic energy scale
$\hbar ^{2}/2ma^{2}$ with lattice spacing $a$. The phonon part of the
kernel, $P_{\varepsilon _{1},\varepsilon _{2},n}=P_{\varepsilon
_{1},\varepsilon _{2},n}^{RPA}+P_{n}^{ph}$ is
\begin{equation}
P_{n}^{ph}=-\frac{g^{2}\Omega ^{2}}{\ \omega _{n}^{2}+\Omega ^{2}}\text{,}
\label{Pph}
\end{equation}%
while in the screened Coulomb part is

\begin{equation}
P_{\varepsilon _{1},\varepsilon _{2},n}^{RPA}=\frac{e^{2}}{\epsilon }%
\int_{\phi =0}^{2\pi }\left\{
\begin{array}{c}
\sqrt{2\left( s-r\cos \phi \right) }+ \\
+\frac{2e^{2}}{\epsilon }\left( 1-\left\vert \omega _{n}\right\vert /\sqrt{%
\omega _{n}^{2}+4\mu \left( s-r\cos \phi \right) }\right)%
\end{array}%
\right\} ^{-1}\text{.}  \label{PRPA}
\end{equation}%
This formula along with the treatment of the neutralizing background is
derived in Appendix B. Here we have used abbreviations%
\begin{eqnarray}
s &=&\varepsilon _{1}+\varepsilon _{2}+2\mu ;  \label{s,p} \\
r &=&2\sqrt{\left( \varepsilon _{1}+\mu \right) \left( \varepsilon _{2}+\mu
\right) }\text{.}  \notag
\end{eqnarray}

To symmetrize the kernel viewed as a matrix, one makes rescaling of the gap
function

\begin{equation}
\eta _{\varepsilon n}=\frac{1}{\sqrt{\omega _{n}^{2}+\varepsilon ^{2}}}%
\Delta _{\varepsilon n}\text{,}  \label{eta}
\end{equation}%
leading to eigenvalue equation%
\begin{equation}
\int_{\varepsilon _{2}=-\mu }^{\Lambda -\mu
}\sum\nolimits_{n_{2}}K_{\varepsilon _{1}n_{1};\varepsilon _{2}n_{2}}\eta
_{\varepsilon _{2}n_{2}}=\eta _{\varepsilon _{1}n_{1}}\text{,}
\label{eigeneta}
\end{equation}%
$\ $where the symmetric matrix is%
\begin{equation}
K_{\varepsilon _{1}n_{1};\varepsilon _{2}n_{2}}=-\frac{mNT}{\pi }\frac{1}{%
\sqrt{\omega _{n_{1}}^{2}+\varepsilon _{1}^{2}}\sqrt{\omega
_{n_{2}}^{2}+\varepsilon _{2}^{2}}}P_{\varepsilon _{1}\varepsilon
_{2},n_{1}-n_{2}}\text{.}  \label{Kdef}
\end{equation}

Critical temperature is obtained when the largest eigenvalue of the matrix $%
K $ is unit. This was done numerically by discretizing variable $\varepsilon
$. The numerical results for the full model are presented in section IV,
however since screening of the STO is very strong we first neglect the
Coulomb repulsion altogether. This allows a significant simplification.

\section{Superconductivity in the LO phonon model}

In this case the theory Eqs.(\ref{He},\ref{Hamiltonian}) has three
parameters (in addition to temperature), the optical phonon frequency $%
\Omega $, the electron-phonon coupling $g$ and chemical potential $\mu $. We
first relate the bare coupling $g$ to the "binding energy $E_{c}$"
conventionally determined in the BCS-BEC crossover studies\cite%
{Randeria2D,Chubukov,BEC2Drev}. Then, since this simplified model will be
applied to the 1UC $FeSe$ on STO, one prefers to parametrize the electron
gas via carrier density $n$ related to the Fermi energy by $\epsilon
_{F}=\pi \hbar ^{2}n/m$ instead of chemical potential $\mu $. Following the
standard practice, $T_{c}$ is found by solving the second Gorkov equation Eq(%
\ref{gapeq}). This is compared with a simpler Kirzhnits approach applied to
the present case in the next section. To simplify the presentation and
without too much loss of generality we take the number of flavors $N=1$.

\subsection{Binding energy}

It is customary\cite{BEC2Drev,Chubukov} to relate the electron - phonon
coupling $g$ to the energy of the bound state $E_{b}\equiv 2E_{c}$ created
by this force in quantum mechanics in vacuum (the two - particle sector of
the multiparticle Hilbert space). We use the binding energy to estimate the
parameter range in which chemical potential $\mu $ approaches the Fermi
energy $\epsilon _{F}$ defined above. In 2D the threshold scattering matrix
element for total energy $E$ at zero momentum obeys the integral
Lippmann-Schwinger equation for scattering amplitude:
\begin{equation}
\Gamma \left( \omega ,\nu ,2E\right) =-V_{\omega -\nu }^{ph}-\frac{1}{2\pi }%
\int_{\rho }V_{\omega -\rho }^{ph}f\left( \rho ,E\right) \Gamma \left( \rho
,\nu ,2E\right) \text{,}  \label{scattering}
\end{equation}%
where

\begin{eqnarray}
f\left( \rho ,E\right) &=&\frac{1}{\left( 2\pi \right) ^{2}}\int_{\mathbf{p}}%
\frac{1}{p^{2}/2m+E+i\rho }\frac{1}{p^{2}/2m+E-i\rho }  \label{fbs} \\
&=&\frac{m}{2\pi }\int_{\varepsilon =E}^{\Lambda }\frac{1}{\varepsilon
^{2}+\rho ^{2}}=\frac{m}{4\left\vert \rho \right\vert }\left( 1-\frac{2}{\pi
}\arctan \frac{E}{\left\vert \rho \right\vert }\right) .  \notag
\end{eqnarray}

The equation Eq.(\ref{scattering}) coincides with the sum of "chain
diagrams" at zero chemical potential in the many - body theory with $\Gamma $
being the "renormalized coupling"\cite{Gorkov}. The bound state (there is
only one such bound state in 2D) with binding energy $2E_{c}$ is found as a
singularity of $\Gamma \left( \omega ,\nu ,2E\right) $. It occurs at energy
for which the matrix of the linear equation Eq.(\ref{scattering}) has zero
eigenvalue, so that the eigenvector $\psi \left( \rho \right) $ obeys%
\begin{equation}
\int_{\rho }\left( 2\pi \delta \left( \omega -\rho \right) +V_{\omega -\rho
}^{ph}f\left( \rho ,E_{c}\right) \right) \psi \left( \rho \right) =0\text{.}
\label{singularity}
\end{equation}%
Changing the variables, $\psi \left( \rho \right) =f\left( \rho ,E\right)
^{-1/2}\eta \left( \rho \right) $, this equation can be presented as the
unit eigenvalue problem
\begin{equation}
\frac{mg^{2}}{2\pi }\int_{\rho }K\left( \omega ,\rho \right) \eta \left(
\rho \right) =\eta \left( \omega \right) \text{,}  \label{eigeneq1}
\end{equation}%
with a symmetric kernel%
\begin{equation}
K\left( \omega ,\rho \right) =\frac{1}{4}\sqrt{\frac{1}{\left\vert \omega
\right\vert }\left( 1-\frac{2}{\pi }\arctan \frac{E_{c}}{\left\vert \omega
\right\vert }\right) \frac{1}{\left\vert \rho \right\vert }\left( 1-\frac{2}{%
\pi }\arctan \frac{E_{c}}{\left\vert \rho \right\vert }\right) }\frac{\Omega
^{2}}{\left( \omega -\rho \right) ^{2}+\Omega ^{2}}\text{.}
\label{Matrixsymbs}
\end{equation}%
It turns out that the unit eigenvalue is the maximal eigenvalue of this
positive definite matrix. The discretized version of Eq.(\ref{eigeneq1}) was
diagonalized numerically. The results are presented in Fig. 1.

\begin{figure}[tbp]
\centering
\includegraphics[width=12cm]{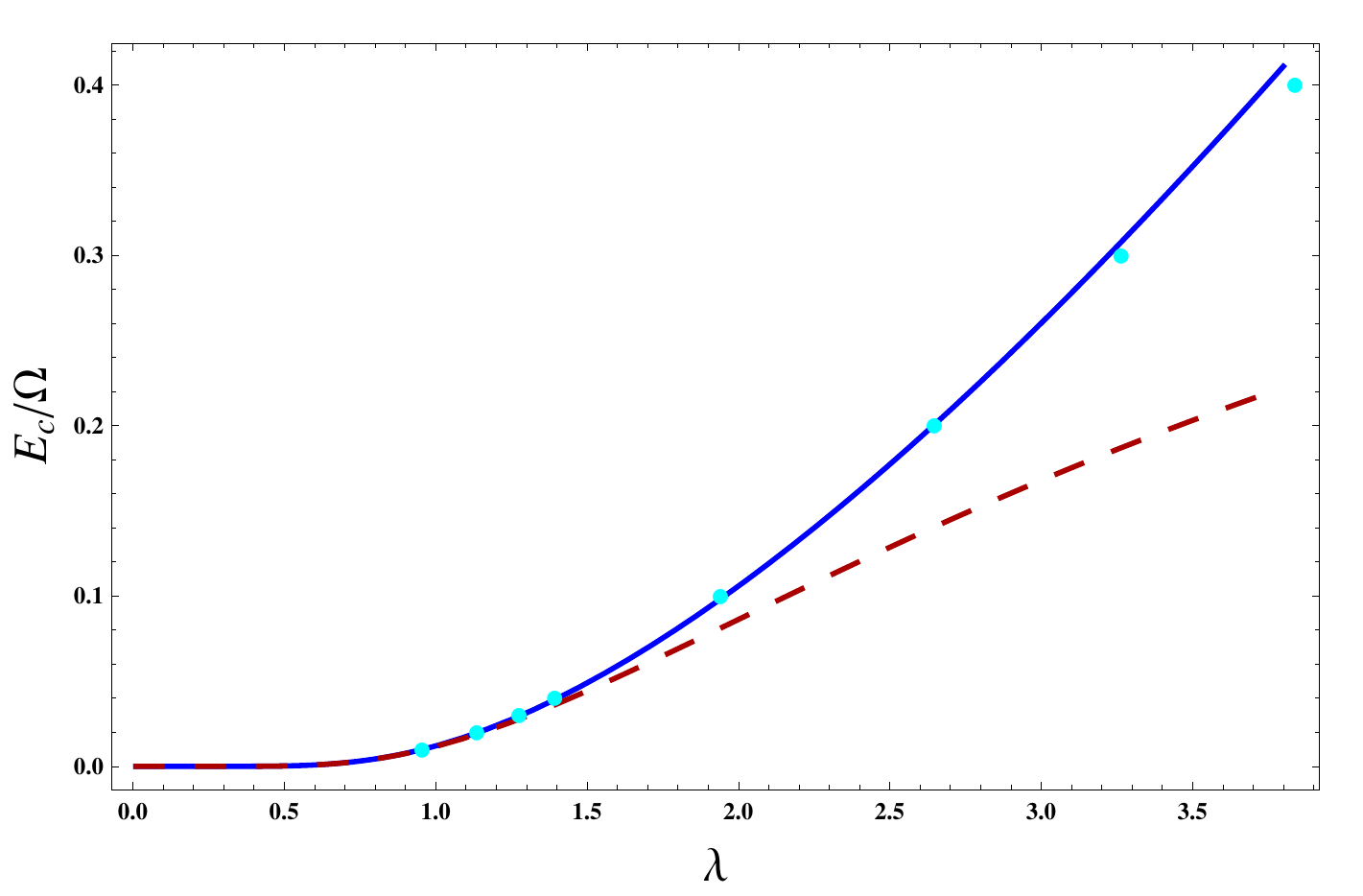} \vspace{-0.5cm}
\caption{The 2D binding energy per electron of two electrons in the bound
state created by the attraction due to LO dispersionless phonon branch with
frequency $\Omega $. The (bare) coupling strength $\protect\lambda $ is in a
wide range $\protect\lambda \sim 0-3.5$. The essential exact dependence
found numerically (dots) is compared with weak coupling (the solid line) and
results obtained using the local model (dashed line).}
\label{1resub}
\end{figure}
Solution found numerically is well fitted by%
\begin{equation}
\frac{2\pi }{mg^{2}}=\frac{1}{\lambda }\approx \frac{1}{2}\sinh ^{-1}\left[
\frac{\Omega \left( \Omega +\pi E_{c}\right) }{\pi E_{c}\left( \Omega
+E_{c}\right) }\right] \text{,}  \label{resultEc}
\end{equation}%
where the 2D dimensionless electron - phonon coupling (per spin) is defined
as $\lambda =\frac{mg^{2}}{2\pi \hbar ^{2}}$. As will be demonstrated in the
following subsections, the interesting range of couplings will obey $%
\epsilon _{F}>>E_{c}$ and thus\cite{Chubukov} we always replace $\mu $ by $%
\epsilon _{F}$.

It has the correct asymptotics at both weak and strong coupling, so that%
\begin{equation}
\frac{E_{c}}{\Omega }=\frac{1}{2\sinh \left[ \frac{2}{\lambda }\right] }%
\left\{ 1-\sinh \left[ \frac{2}{\lambda }\right] +\sqrt{\left( 1-\sinh \left[
\frac{2}{\lambda }\right] \right) ^{2}+\frac{4}{\pi }\sinh \left[ \frac{2}{%
\lambda }\right] }\right\} \text{.}  \label{resEc2}
\end{equation}%
At weak coupling
\begin{equation}
E_{c}/\Omega =\frac{2}{\pi }e^{-2/\lambda }<<1  \label{Chubukov}
\end{equation}%
and hence one can use a local "instantaneous" electron-phonon interaction
model, with Eq.(\ref{Pph}) approximated by%
\begin{equation}
P_{n}^{ph}=-\frac{g^{2}\Omega ^{2}}{\ \omega _{n}^{2}+\Omega ^{2}}\approx
-g^{2}\theta \left( \left\vert \omega _{n}\right\vert -\Omega \right) \text{,%
}  \label{instant}
\end{equation}%
to describe this limit. In the instantaneous model the electron - phonon
interaction is assumed to vanish on the scale of $\Omega $. Therefore in
this approximation for $\epsilon _{F}<<\Omega $ all the integrations can be
cut off at this scale intercepting the larger cutoff $\Lambda $. The results
for $E_{c}$ are consistent with BEC literature\cite{Chubukov}, see dashed
line in Fig. 1. Note that the dimensionless pre-exponential factor in Eq.(%
\ref{Chubukov}) is determined to be $\frac{2}{\pi }$.

\subsection{The energy independence of the gap function}

The equation Eq.(\ref{inteq}) in the limit $e^{2}\rightarrow 0$ is:%
\begin{equation}
\frac{g^{2}mT}{2\pi }\sum\nolimits_{n_{2}}\int_{\varepsilon _{2}=-\epsilon
_{F}}^{\Lambda -\epsilon _{F}}\frac{1}{\omega _{n_{2}}^{2}+\varepsilon
_{2}^{2}}\frac{\Omega ^{2}}{\left( \omega _{n_{1}}-\omega _{n_{2}}\right)
^{2}+\Omega ^{2}}\Delta _{\varepsilon _{2}n_{2}}=\Delta _{\varepsilon
_{1}n_{1}}\text{.}  \label{gapphon}
\end{equation}%
Since the left hand side of the equation is independent of $\varepsilon _{2}$%
, the gap function is independent of energy: $\Delta _{\varepsilon n}=\Delta
_{n}$. Substituting this, one gets a one dimensional integral equation%
\begin{eqnarray}
&&\lambda T\sum\nolimits_{n_{2}}\frac{\Omega ^{2}}{\left( \omega
_{n_{1}}-\omega _{n_{2}}\right) ^{2}+\Omega ^{2}}\Delta
_{n_{2}}\int_{\varepsilon _{2}=-\epsilon _{F}}^{\Lambda -\epsilon _{F}}\frac{%
1}{\omega _{n_{2}}^{2}+\varepsilon _{2}^{2}}  \label{gapeqph} \\
&=&\lambda \sum\nolimits_{n_{2}}\frac{\Omega ^{2}f\left( \omega
_{n_{2}}\right) }{\left( \omega _{n_{1}}-\omega _{n_{2}}\right) ^{2}+\Omega
^{2}}\Delta _{n_{2}}=\Delta _{n_{1}}\text{,}  \notag
\end{eqnarray}%
where the integral is%
\begin{equation}
f\left( \omega \right) =\frac{T}{\left\vert \omega \right\vert }\left(
\arctan \frac{\Lambda -\epsilon _{F}}{\left\vert \omega \right\vert }%
+\arctan \left[ \frac{\epsilon _{F}}{\left\vert \omega \right\vert }\right]
\right) \text{.}  \label{fdef}
\end{equation}%
Changing of variables $\eta _{n}=\sqrt{f\left( \omega _{n}\right) }\Delta
_{n}$, makes the kernel matrix of the integral equation,
\begin{equation}
\sum\nolimits_{n_{2}}K_{n_{1}n_{2}}\left( T\right) \eta _{n_{2}}=\eta
_{n_{1}}\text{,}  \label{eigeneq}
\end{equation}%
symmetric,%
\begin{equation}
K_{n_{1}n_{2}}\left( T\right) =\lambda \frac{\sqrt{f\left( \omega
_{n_{1}}\right) f\left( \omega _{n_{2}}\right) }\Omega ^{2}}{\left( \omega
_{n_{1}}-\omega _{n_{2}}\right) ^{2}+\Omega ^{2}}\text{.}  \label{Kphon}
\end{equation}

\subsection{Numerical procedure and results}

The eigenvalue equation Eq(\ref{eigeneq}) is solved numerically by
diagonalizing sufficiently large matrix $K_{n_{1}n_{2}}\left( T\right) $.
The index $-N_{\omega }/2<n<N_{\omega }/2$ with the value $N_{\omega }=256$
used. At this value of $N_{\omega }$ the results are already independent of
the UV cutoff $\Lambda $. The critical temperature for given $\lambda $, $%
\epsilon _{F}$ and $\Omega $ is determined from the requirement that the
largest eigenvalue of $K\left( T\right) $ is $1$. The results presented as
functions of $\epsilon _{F}$ in Fig. 2 in whole range of $\epsilon _{F}$ and
Fig. 3 for $\epsilon _{F}<\Omega $.
\begin{figure}[tbp]
\centering
\includegraphics[width=12cm]{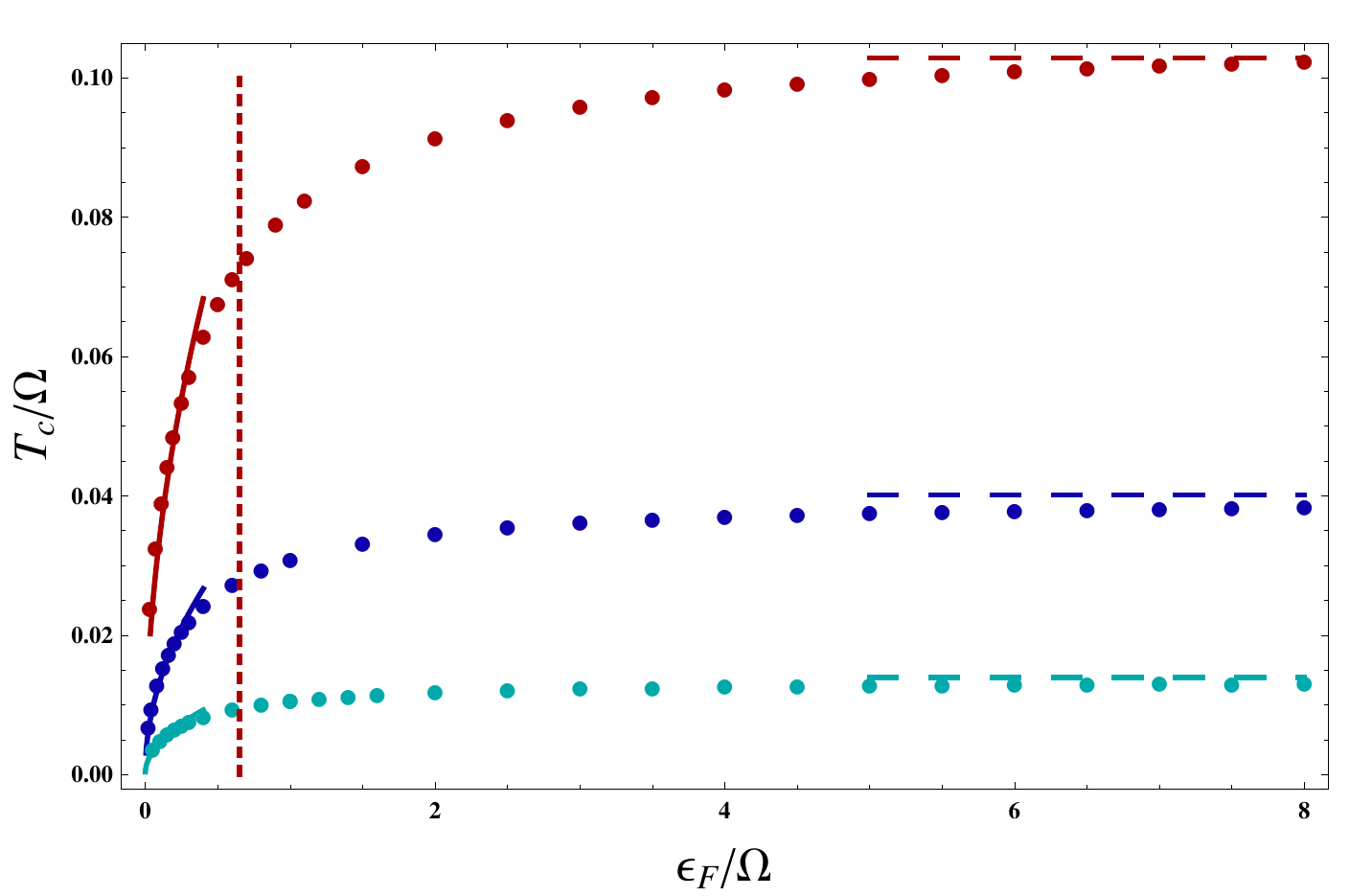} \vspace{-0.5cm}
\caption{The critical temperature of a 2DEG - LO phonon superconductor (the
Coulomb repulsion is assumed to screened out by the substrate). $T_{c}$ in
units of the phonon frequency $\Omega $ is given as a function of the Fermi
energy in whale range of $\protect\epsilon _{F}/\Omega ,$for the
dimensionless electron-phonon coupling (from top to bottom): $\protect%
\lambda =0.5,0.34,0.25$. The adiabatic\ (BCS) limit is a dashed line. Solid
line is result of local theory.}
\end{figure}

\begin{figure}[tbp]
\centering
\includegraphics[width=12cm]{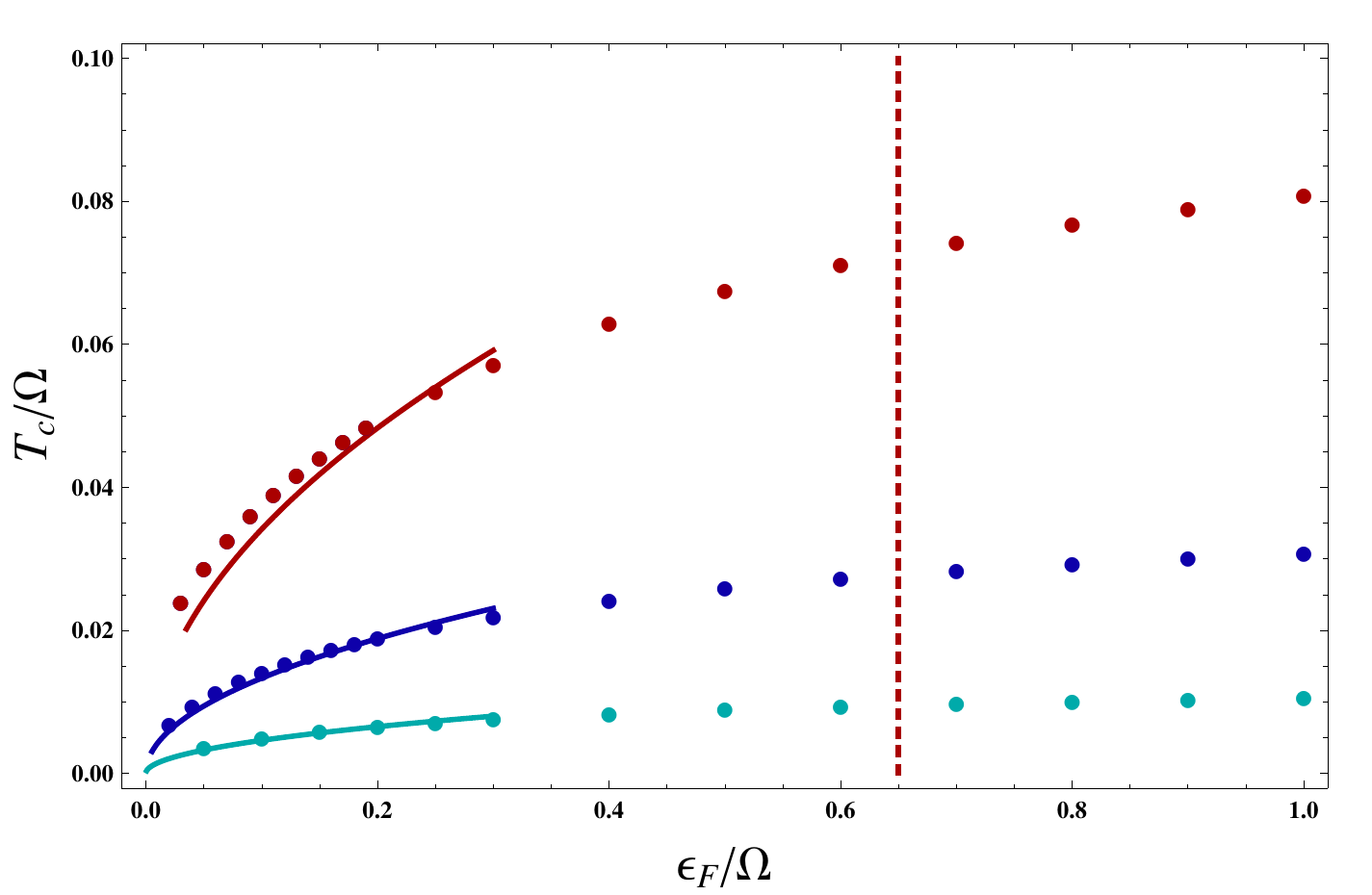} \vspace{-0.5cm}
\caption{ The critical temperature of a 2DEG - LO phonon superconductor in
the low temperatures range in units of the phonon frequency $\Omega $ for $%
\protect\lambda =0.5,0.34,0.25.$ Solid line is the result of the local
theory. }
\end{figure}

\subsection{Adiabatic and nonadiabatic (local interaction model) limits}

In the strongly adiabatic situation, $\epsilon _{F}>>\Omega $, one can take
the $\epsilon _{F}\rightarrow \infty $ limit in which the matrix simplifies,
$f\left( \omega \right) \approx \frac{\pi T}{\left\vert \omega \right\vert }%
, $

\begin{equation}
K_{n_{1}n_{2}}^{BCS}\left( T\right) =\frac{\lambda }{\sqrt{\left\vert
n_{1}+1/2\right\vert \left\vert n_{2}+1/2\right\vert }\left( \left( 2\pi
\frac{T}{\Omega }\left( n_{1}-n_{2}\right) \right) ^{2}+1\right) }\text{.}
\label{KBCS}
\end{equation}%
This can be fitted by the phenomenological McMillan like formula (dashed
lines in Fig.2),

\begin{equation}
T_{c}^{adiab}\left( \lambda \right) \approx 0.75\text{ }\Omega \exp \left[ -%
\frac{1}{\lambda }\right] \text{.}  \label{McMillan}
\end{equation}%
In the opposite strongly non-adiabatic limit, $E_{c}<<\epsilon _{F}<<\Omega $%
, the local model defined in subsection A can be used. The gap equation Eq.(%
\ref{gapeqph}) for frequency independent $\Delta _{n}=\Delta $ simplifies
into

\begin{equation}
\lambda \sum\nolimits_{n_{2}=-\Omega /\left( 2\pi T_{c}\right) }^{\Omega
/\left( 2\pi T_{c}\right) }f\left( \omega _{n_{2}}\right) \Delta =\Delta
\text{.}  \label{gaplocal}
\end{equation}%
The solution exists for%
\begin{equation}
\lambda T_{c}\sum\nolimits_{n=-\Omega /\left( 2\pi T_{c}\right) }^{\Omega
/\left( 2\pi T_{c}\right) }\frac{1}{\left\vert \omega _{n}\right\vert }%
\left( \frac{\pi }{2}+\arctan \left[ \frac{\epsilon _{F}}{\left\vert \omega
_{n}\right\vert }\right] \right) =1  \label{gaploc1}
\end{equation}%
At low temperatures the sum can be approximated by an integral%
\begin{equation}
\frac{\lambda }{\pi }\int_{\omega =\pi T_{c}}^{\Omega }\frac{1}{\omega }%
\left( \frac{\pi }{2}+\arctan \left[ \frac{\epsilon _{F}}{\omega }\right]
\right) =1\text{,}  \label{integralTc}
\end{equation}%
one gets the formula

\begin{equation}
T_{c}^{local}\left( \lambda \right) =\sqrt{E_{c}\left( \lambda \right)
\epsilon _{F}}=\sqrt{\frac{2\Omega \epsilon _{F}}{\pi }}\exp \left[ -\frac{1%
}{\lambda }\right] \text{.}  \label{McMillanBEC}
\end{equation}%
The curves are given in Fig.3 (dashed lines) and compares well with the
simulated result (circles) for $\lambda =0.5,0.34,0.25$ (from top to bottom).

There exists an alternative approach to such calculations (beyond the
Gaussian approximation adopted here), see \cite{Melik} in which the
correlator at zero chemical potential is subtracted. We don't use it, but
very recently Chubukov et al found \cite{Chubukov} that for the local
instantaneous model results are identical. It is instructive to compare the
direct numerical simulation with a simpler approximate semi - analytic
Kirzhnits method that is applied to the model in the following Section.

\section{Comparison with the Kirzhnits Ansatz}

\subsection{Application of the Kirzhnits method to LO phonon model}

Integral equations in general case Eqs.(\ref{eigeneq}) are very complicated
and typically approximated by simpler one dimensional integral equations. It
was first proposed long time ago by Kirzhnits\cite{Kirzhnits,Takada} and
later developed for the dielectric function approach to novel
superconductors \cite{Klimin14}. In this section the units of $\hbar
=m=\Omega =1$ and physical frequency (not Matsubara) is used. Spectral
representation of the dispersionless optical phonon contribution to inverse
dielectric constant is:

\begin{equation}
\sigma \left( k,E\right) =\frac{\epsilon }{e^{2}}\lambda k\delta \left(
1-E^{2}\right) \text{.}  \label{sigmaK}
\end{equation}%
The gap equation for the quantity characterizing the anomalous average $%
F_{p} $ defined by Kirzhnits\cite{Kirzhnits} reads,%
\begin{equation}
\Phi \left( p\right) =-\frac{e^{2}}{2\pi \epsilon }\int_{\mathbf{k}}\frac{%
B\left( \varepsilon _{k}\right) }{\left\vert \mathbf{p-k}\right\vert }\left(
1-2\int_{E=0}^{\Lambda }\frac{\sigma \left( \left\vert \mathbf{p-k}%
\right\vert ,E\right) }{E+\left\vert \varepsilon _{k}\right\vert +\left\vert
\varepsilon _{p}\right\vert }\right) \Phi \left( k\right) \text{,}
\label{gapeqK}
\end{equation}%
where

\begin{equation}
B\left( \varepsilon _{k}\right) =\frac{\tanh \left( \varepsilon
_{k}/2T_{c}\right) }{2\varepsilon _{k}}\text{.}  \label{BR}
\end{equation}%
Substituting Eq.(\ref{BR}) into Eq.(\ref{gapeqK}), and transforming the
variable $k$ to the energy, one obtains:
\begin{equation}
\Phi \left( p\right) =\lambda \int_{\varepsilon _{k}=-\epsilon
_{F}}^{\Lambda -\epsilon _{F}}\frac{B\left( \varepsilon _{k}\right) }{%
1+\left\vert \varepsilon _{k}\right\vert +\left\vert \varepsilon
_{p}\right\vert }\Phi \left( k\right) \text{.}  \label{gapeqK1}
\end{equation}%
Symmetrization of the kernel, $\Phi \left( p\right) =\sqrt{B\left(
\varepsilon _{p}\right) }\eta _{p}$, one obtains:
\begin{equation}
\lambda \int_{\varepsilon _{2}=-\epsilon _{F}}^{\Lambda -\epsilon _{F}}\frac{%
\sqrt{B\left( \varepsilon _{1}\right) B\left( \varepsilon _{2}\right) }}{%
1+\left\vert \varepsilon _{1}\right\vert +\left\vert \varepsilon
_{2}\right\vert }\eta _{2}=\eta _{1}\text{.}  \label{symK}
\end{equation}%
This is solved numerically for $\epsilon _{F}=0.5,1,5\Omega $ and $\lambda
=0.5$ with the ultraviolet cutoff $\Lambda =15\Omega $ in the upper limit of
integral in Eq.(\ref{symK}) with number of values of energy $N_{\varepsilon
}=4000$, so that the step is smaller than $\left( \epsilon _{F}+\Lambda
\right) /N_{\varepsilon }\sim 10^{-2}$. The results are presented in Fig.4,5
as starts.

\begin{figure}[tbp]
\centering
\includegraphics[width=12cm]{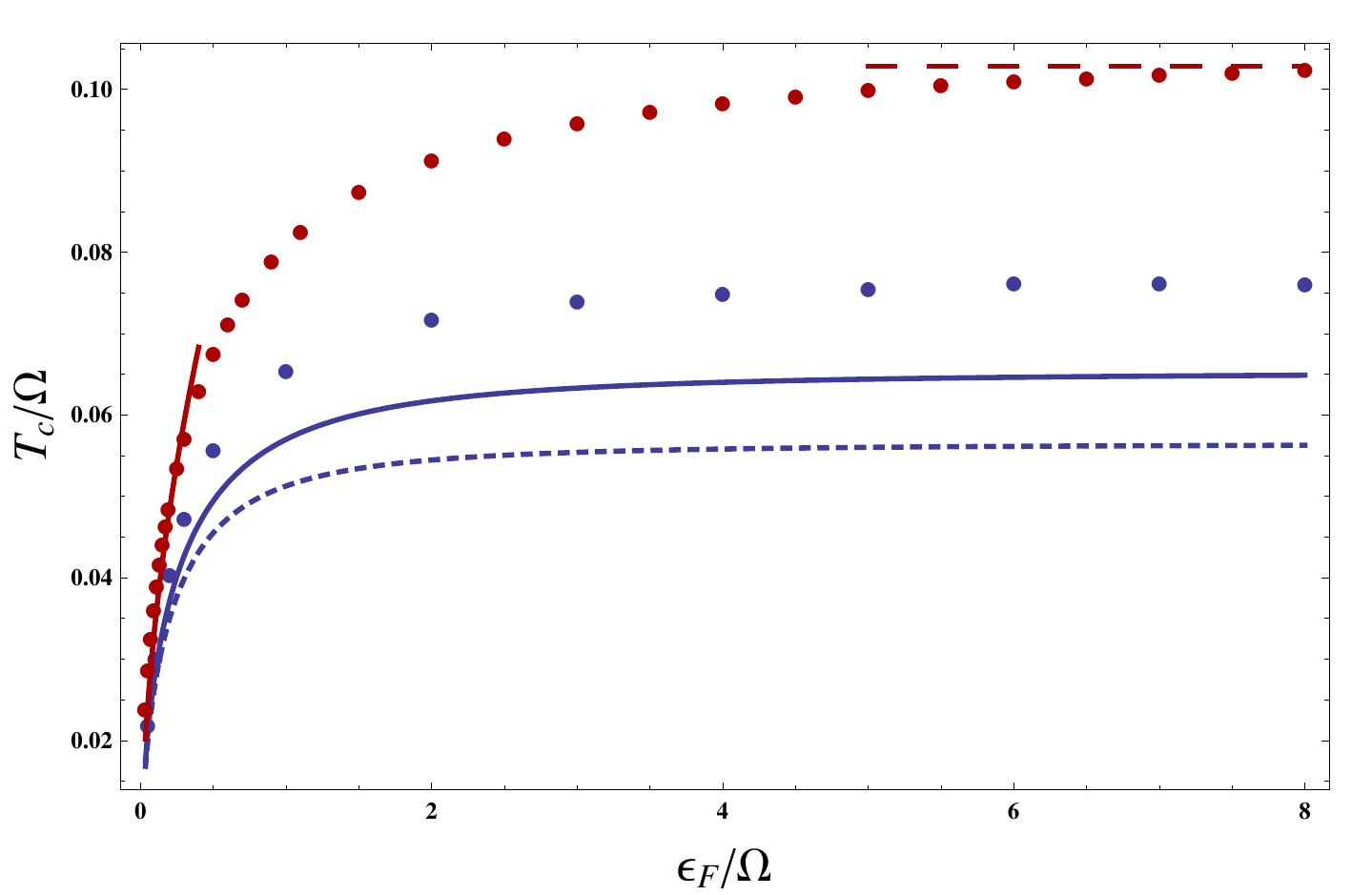} \vspace{-0.5cm}
\caption{Comparison with the critical temperature of the Kirzhnits Ansatz
approximation for a wide range of Fermi energies. The brown dots are the
same as in Fig.2 for $\protect\lambda =0.5$ while the solid line there is
the result of the instantaneous (local) theory. The Kirzhnits approximation $%
T_{c}$ calculated numerically is given by blue dots, while the dashed \ and
solid blue lines are the weak coupling approximation analytic results at
leading and the next to leading order respectively.}
\end{figure}

\begin{figure}[tbp]
\centering
\includegraphics[width=12cm]{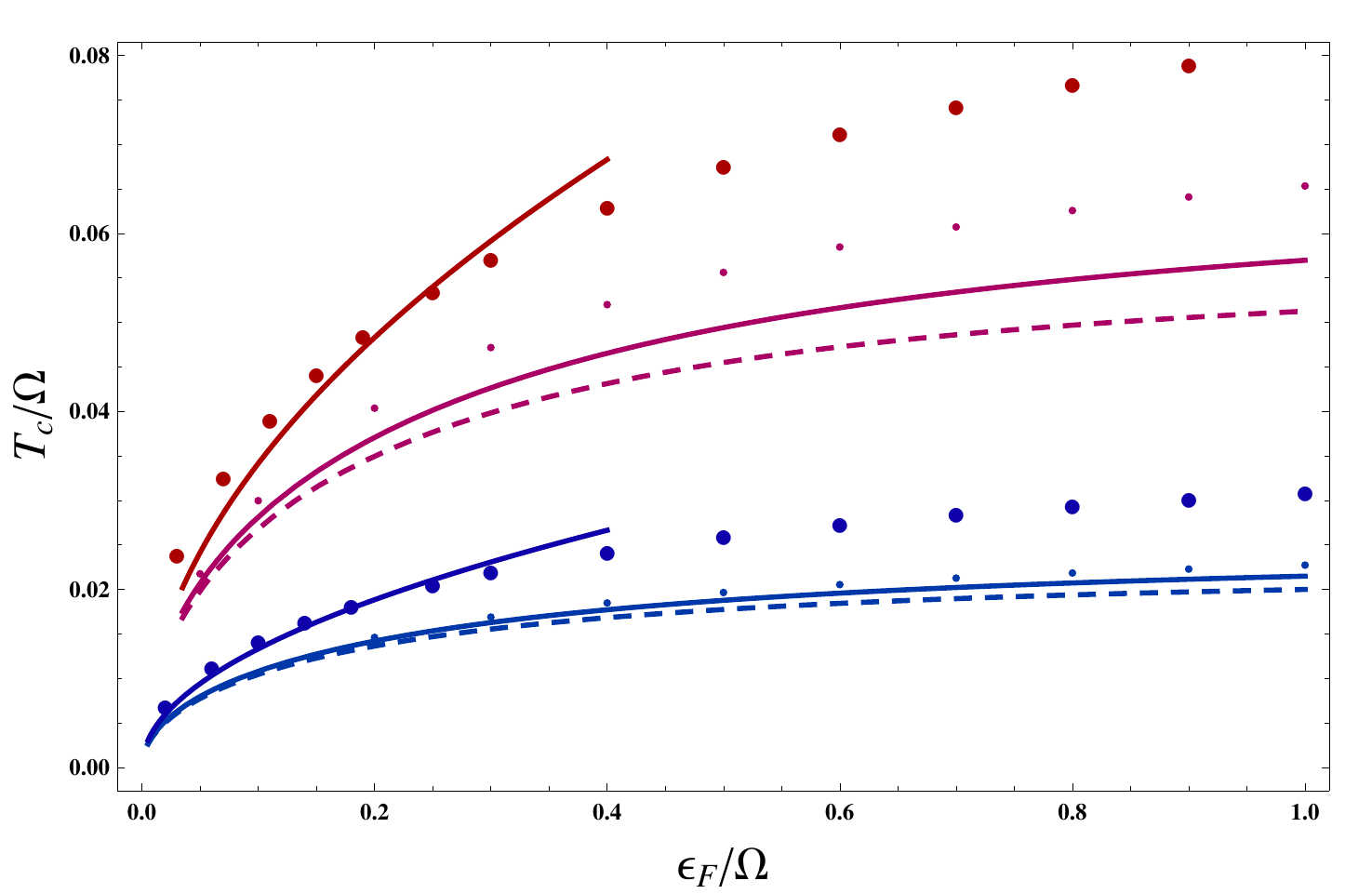} \vspace{-0.5cm}
\caption{Comparison with the critical temperature of the Kirzhnits Ansatz
approximation for a small Fermi energies. The brown dots are the result of
numerical solution of the gap equation and are the same as in Fig.2 for $%
\protect\lambda =0.5,$and $0.34$ where solid line is the result of
instantaneous theory. The Kirzhnits approximation $T_{c}$ calculated
numerically is given by blue dots, while the dashed \ and solid blue lines
are the weak coupling approximation analytic results at leading and the next
to leading order respectively. }
\end{figure}
It is possible to obtain a closed analytic expression only at weak coupling.

\subsection{Weak coupling}

At small coupling the critical temperature can be estimated analytically
using the asymptotic theory due to Zubarev\cite{Zubarev}:
\begin{equation}
T_{c}=\frac{2}{\pi }e^{\gamma _{E}}\epsilon _{F}e^{-\frac{1}{\lambda }%
}e^{\zeta \left( \epsilon _{F},\lambda \right) }\text{,}  \label{B5}
\end{equation}%
where%
\begin{equation}
\zeta \left( \epsilon _{F},\lambda \right) =\int_{\varepsilon =-\epsilon
_{F}}^{\infty }\frac{1}{2\left\vert \varepsilon \right\vert }\left( \frac{%
\phi _{\varepsilon }}{1+\left\vert \varepsilon \right\vert }-\Theta \left(
\epsilon _{F}-\varepsilon \right) \right) \text{.}  \label{B6}
\end{equation}%
Equation determining $\phi _{\varepsilon }\equiv \eta _{\varepsilon }/\eta
_{\varepsilon =0}$ for small temperatures is approximated in our case by:

\begin{equation}
\phi _{\varepsilon }-\frac{\lambda \left\vert \varepsilon \right\vert }{%
2\left( 1+\left\vert \varepsilon \right\vert \right) }\int_{\varepsilon
^{\prime }=-\epsilon _{F}}^{\infty }\frac{\phi _{\varepsilon ^{\prime }}}{%
\left( 1+\left\vert \varepsilon \right\vert +\left\vert \varepsilon ^{\prime
}\right\vert \right) \left( 1+\left\vert \varepsilon ^{\prime }\right\vert
\right) }=\frac{1}{1+\left\vert \varepsilon \right\vert }.  \label{B3}
\end{equation}%
This is solved iteratively to second order, $\phi _{\varepsilon }=\phi
_{\varepsilon }^{\left( 0\right) }+\lambda \phi _{\varepsilon }^{\left(
1\right) }$,%
\begin{eqnarray}
\phi _{\varepsilon } &=&\frac{1}{1+\left\vert \varepsilon \right\vert }%
+\lambda \phi _{\varepsilon }^{\left( 1\right) }  \label{B4} \\
\phi _{\varepsilon }^{\left( 1\right) } &=&\frac{1}{2\left( 1+\left\vert
\varepsilon \right\vert \right) }\left\{ \frac{1+2\epsilon _{F}}{\left(
1+\epsilon _{F}\right) }-\frac{1}{\left\vert \varepsilon \right\vert }\log
\frac{\left( 1+\left\vert \varepsilon \right\vert \right) ^{2}\left(
1+\epsilon _{F}\right) }{1+\left\vert \varepsilon \right\vert +\epsilon _{F}}%
\right\} \text{.}  \notag
\end{eqnarray}%
Substituting this into Eq. (\ref{B6}) one obtains,%
\begin{eqnarray}
\zeta \left( \epsilon _{F},\lambda \right) &=&\zeta ^{0}\left( \epsilon
_{F}\right) +\lambda \zeta ^{1}\left( \epsilon _{F}\right) +O\left( \lambda
^{2}\right)  \label{B7} \\
\zeta ^{\left( 0\right) }\left( \epsilon _{F}\right) &=&\int_{\varepsilon
=-\epsilon _{F}}^{\infty }\frac{1}{2\left\vert \varepsilon \right\vert }%
\left\{ \frac{1}{\left( 1+\left\vert \varepsilon \right\vert \right) ^{2}}%
-\Theta \left( \epsilon _{F}-\varepsilon \right) \right\} =-\frac{1}{2}%
\left\{ \frac{1+2\epsilon _{F}}{1+\epsilon _{F}}+\log \left[ \epsilon
_{F}\left( 1+\epsilon _{F}\right) \right] \right\} \text{.}  \notag
\end{eqnarray}%
The second correction,%
\begin{equation}
\zeta ^{\left( 1\right) }\left( \epsilon _{F}\right) =\int_{\varepsilon
=-\epsilon _{F}}^{\infty }\frac{\phi _{\varepsilon }^{\left( 1\right) }}{%
2\left\vert \varepsilon \right\vert \left( 1+\left\vert \varepsilon
\right\vert \right) }\text{,}  \label{B8}
\end{equation}%
still can be calculated analytically via hypergeometric function, but is
cumbersome. It is regular and for $\lambda =0.5$ corrects the analytic
result shown in Fig.4,5 as a dotted line into the one (solid line) closer to
numerical solution. The formula works better for nonadiabatic regime, Fig.5,
than in the adiabatic limit, Fig.4.

The approximate formula neglecting the second order correction in the
adiabatic regime, $\epsilon _{F}>\Omega $,is
\begin{equation}
T_{c}=\frac{2}{\pi }e^{\gamma _{E}}\epsilon _{F}e^{-\frac{1}{\lambda }}\exp %
\left[ -1-\log \left[ \epsilon _{F}\right] \right] =\frac{2}{\pi }e^{\gamma
_{E}-1}\Omega e^{-\frac{1}{\lambda }}\approx 0.41\Omega \text{ }e^{-\frac{1}{%
\lambda }}\text{.}  \label{adiabatK}
\end{equation}%
The coefficient is significantly smaller than the fit to the numerical
solution, Eq.(\ref{McMillan}). In the opposite nonadiabatic limit
\begin{equation}
T_{c}=\frac{2}{\pi }e^{\gamma _{E}}\epsilon _{F}e^{-\frac{1}{\lambda }}\exp %
\left[ -\frac{1}{2}\left\{ 1+\log \left[ \epsilon _{F}\right] \right\} %
\right] =\frac{2}{\pi }e^{\gamma _{E}-1/2}\sqrt{\Omega \epsilon _{F}}e^{-%
\frac{1}{\lambda }}\approx 0.69\sqrt{\Omega \epsilon _{F}}\text{exp}\left[ -%
\frac{1}{\lambda }\right] \text{.}  \label{nonadiabat}
\end{equation}

To conclude the critical temperature in the Kirzhnits approach is generally
underestimated by 30\% in adiabatic limit and is precise in the nonadiabatic
limit. Within the range of applicability the general tendency is correct.
Next we tackle a more complicated model incorporating the effect of the
screened Coulomb repulsion.

\section{The effect of the Coulomb repulsion}

The eigenvalue equation Eq.(\ref{eigeneta}) with the kernel including the
RPA dynamically screened Coulomb repulsion, Eq.(\ref{PRPA}), is solved
numerically by diagonalizing sufficiently large matrix $K_{n_{1}\varepsilon
_{1,}n_{2}\varepsilon _{2}}\left( T\right) $. In the presence of moderately
screened Coulomb repulsion,\ to describe the ALO/STO interfaces, the
chemical potential is practically equal to the Fermi energy $\epsilon _{F}$.

The integral over the angle $\phi $ in Eq.(\ref{PRPA}) was performed
numerically ($720$ subdivisions). The neutralizing background was subtracted
(the screening is dynamic, so that the interaction is generally still long
range, see Appendix B). The Matsubara index is in the range $-N_{\omega
}/2<n<N_{\omega }/2$ with the value $N_{\omega }=16$ used. The energy cutoff
was in the range $\Lambda =3\epsilon _{F}$ (for nonadiabatic values $%
\epsilon _{F}=0.5$, $1$) and up to $\Lambda =15\epsilon _{F}$ in the
adiabatic regime. Number of values of energy $N_{\varepsilon }=256$, so that
the step is smaller than $\left( \epsilon _{F}+\Lambda \right)
/N_{\varepsilon }\sim 2.4\times 10^{-3}$. Convergence was checked against
higher values of $\Lambda ,N_{e}$ and $N_{\omega }$.

The critical temperature for given $\lambda $, $m$, $\epsilon _{F}$ and $%
\Omega $ is determined from the requirement that the largest eigenvalue of $%
K\left( T\right) $ is $1$. The use units in which $\hbar =\Omega =m=1$. In
these units the Coulomb couplings become%
\begin{equation}
\alpha =\frac{e^{2}m^{1/2}}{\epsilon \Omega ^{1/2}\hbar }\text{.}
\label{alfa}
\end{equation}%
For $\Omega =1000K,m=m_{e},\epsilon =3000$ one gets $\alpha =6\times 10^{-3}$%
. The results presented in Fig.6 in the Coulomb coupling range $5\times
10^{-3}-7\times 10^{-2}$ sufficient for our purposes. One clearly observes
the Coulomb suppression that is not homogeneous in $\epsilon _{F}$. At $%
\epsilon _{F}$ comparable with $\Omega $ or slightly smaller (the smallest
simulated value is $\epsilon _{F}=0.5\Omega $) one observes that at larger $%
\alpha $ an approach to the BCS limit is slower.

\begin{figure}[tbp]
\centering
\includegraphics[width=12cm]{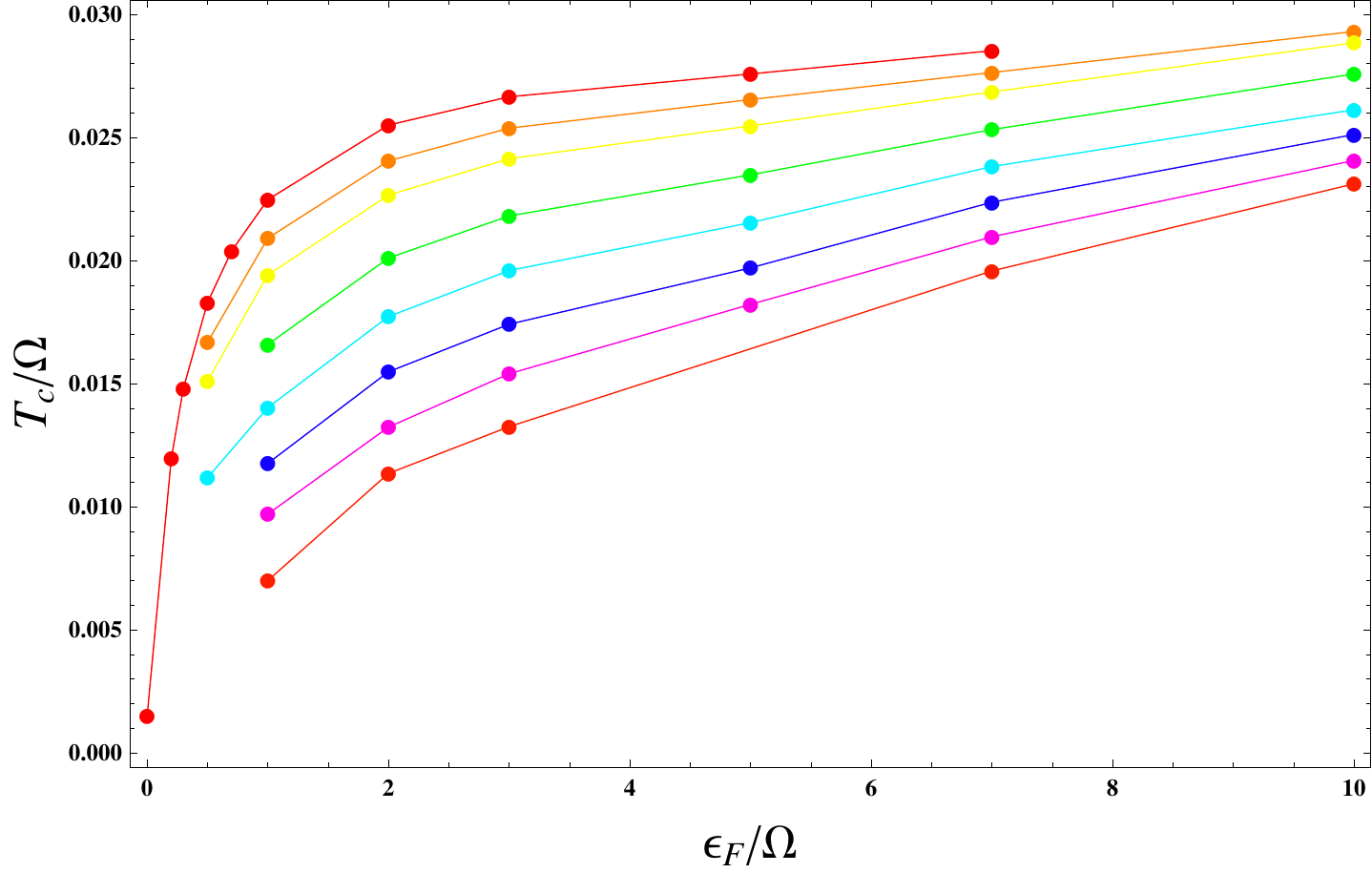} \vspace{-0.5cm}
\caption{Suppression of the critical temperature of a 2DEG phonon
superconductor the RPA screened Coulomb repulsion. $T_{c}$ in units of the
phonon frequency $\Omega $ for $\protect\lambda =0.32$ is given as a
function of the chemical potential for the following dimensionless effective
Coulomb repulsion strength $\protect\alpha $ defined in Eq.(\protect\ref%
{alfa}). From top to bottom: $\protect\alpha =0$ (the phonon model, red
dots), $\protect\alpha =5\cdot 10^{-3}$ (brown dots), $\protect\alpha %
=10^{-2}$ (yellow), $\protect\alpha =2\cdot 10^{-2}$ (green), $\protect%
\alpha =3\cdot 10^{-2}$ (blue), $\protect\alpha =4\cdot 10^{-2}$ (violet),$%
\protect\alpha =5\cdot 10^{-2}$ (pink), $\protect\alpha =6\cdot 10^{-2}$
(dark red). The curves are well approximated by the interpolating formula,
Eqs.(\protect\ref{fit_alfa}).}
\end{figure}

A reasonable interpolation formula for all the values is:%
\begin{equation}
T_{c}\left( \Omega ,\epsilon _{F},\lambda \right) =0.8\text{ }\Omega \exp %
\left[ -\frac{2}{\lambda -1.2\alpha }\frac{\Omega +3\epsilon _{F}}{\Omega
+6\epsilon _{F}}\right] \text{.}  \label{fit_alfa}
\end{equation}%
We use this formula to discuss the interface superconductivity in the next
Section.

\section{Application to superconductivity in 1D$FeSe$/STO substrate and
related materials}

\subsection{1UC$FeSe$/STO}

Based on experiments described in the introduction, the following parameters
should be used in the simple LO model of Section IV. \ The phonon frequency
was estimated by ARPES\cite{Lee14} to in the in $\Omega =80-100meV$ range
and by the ultrafast dynamics\cite{ultrafast16} to be $\Omega =106meV$. The
dimensionless electron-phonon coupling constant was estimated (using a model
with a flat phonon spectrum) from the intensity ratios in ARPES \cite{Lee14}
to be $\lambda =0.5,$ consistent with $\lambda =0.48$ from the ultrafast
dynamics\cite{ultrafast16}. The critical temperature estimates were rather
scattered and dependent on the method. While the critical temperature
deduced from the gap in tunneling is $T_{c}=70K$, magnetization experiments%
\cite{Wang14} indicate that $T_{c}=85K$ and the ultrafast\cite{ultrafast16}
dynamics gives $T_{c}=68K$. The temperature was directly measured in
transport\cite{Jia15} to be $100K$. The Fermi surface\cite{Lee14} for the
electron pockets is located at $\epsilon _{F}=60meV$.

In the simplified model of Section II (neglecting completely the Coulomb
repulsion due to the huge dielectric constant of STO) the only parameters
determining $T_{c}$ are $\lambda ,\Omega $ and $\epsilon _{F}$. This is
presented in Figs. 2,3. Taking $\Omega =100mev$, $\epsilon _{F}=60mev$ one
obtains for $\lambda =0.5$ ,$T_{c}=77K\,$, see the dotted line in Figs. 2,3.
This is within the experimentally possible range. The 2UC FeSe/STO already
has three pockets and resembles the parent material more than 1UC$FeSe$/STO.

\subsection{Interface superconductivity in LAO/STO}

In this case the dielectric constant is one order of magnitude smaller ($%
\epsilon _{0}=186$ on the STO side and $\epsilon _{0}=24$ on the LAO side,
see ref.\cite{Klimin12} where accurate electrostatics was considered) than
in 1UC$FeSe$/STO. Consequently the Coulomb repulsion cannot be neglected,
especially in view of very low $T_{c}\sim 0.2K$. Therefore we have to use
the full model of Section IV. In this case one takes $N=1$ and effective
mass $m=1.65m_{e}$ (where $m_{e}$ is the electron mass in vacuum). Recently%
\cite{Mannhart15} the electron-phonon coupling and chemical potential were
measured by tunneling from the underdoped to the overdoped region. Generally
in the underdoped region the chemical potential rises linearly with the gate
voltage $V_{g}$, $\epsilon _{F}\left( V_{g}\right) =\mu _{0}\left( 1+\eta
V_{g}\right) ,$ with the slope $\eta =1.8\times 10^{-3}V^{-1}$ and is
saturated in the overdoped region at value $\mu _{0}=30meV$. The
electron-phonon coupling apparently decreases very slowly, $\lambda =\lambda
_{0}\left( 1-\gamma V_{g}\right) $, where $\lambda _{0}=0.28$ is the undoped
value, and $\gamma =1.1\times 10^{-4}V^{-1}$ is the slope. Our approximate
formula Eq.(\ref{fit_alfa}) in this case gives the dependence
\begin{equation}
T_{c}\left( V_{g}\right) =0.8\text{ }\Omega \exp \left[ -\frac{2}{\lambda
_{0}\left( 1-\gamma V_{g}\right) -1.2\alpha }\frac{1+3\mu _{0}\left( 1+\eta
V_{g}\right) }{1+6\mu _{0}\left( 1+\eta V_{g}\right) }\right] \text{.}
\label{voltage}
\end{equation}

Taking a measured value for the LO4 mode $\Omega =99.3meV$, lets us estimate
the Coulomb repulsion constant as%
\begin{equation*}
\alpha =\frac{e^{2}m^{1/2}}{\epsilon _{eff}\Omega ^{1/2}\hbar }=0.09
\end{equation*}%
for $\epsilon _{eff}=200$. Substituting these values one obtains the fit to
experimental values of ref.\cite{Mannhart15}, see Fig.7.

\begin{figure}[tbp]
\centering
\includegraphics[width=12cm]{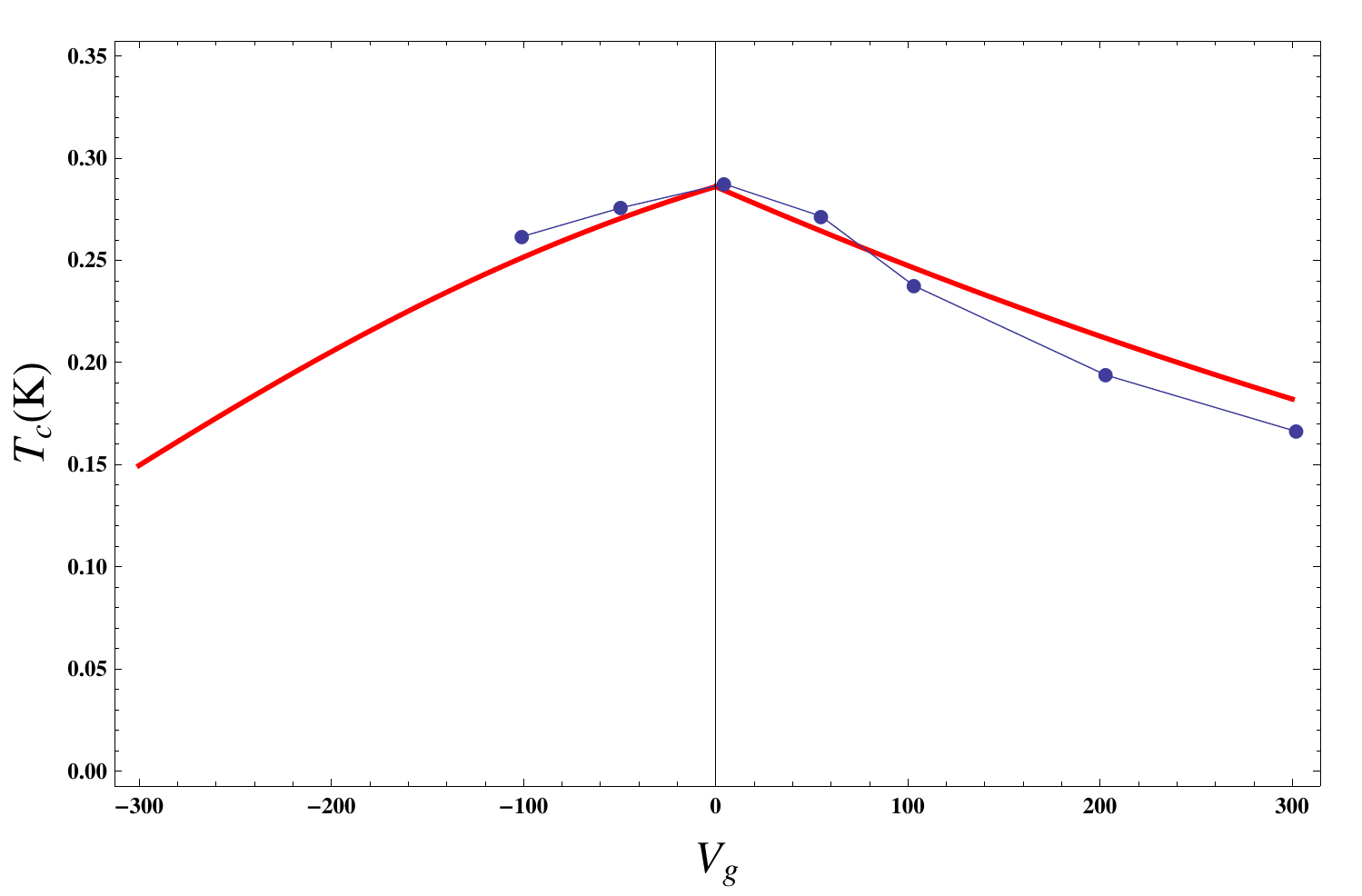} \vspace{-0.5cm}
\caption{Tc as a function on gate voltage $V_g$.}
\end{figure}

Qualitatively there are two conflicting tendencies at play. The reduction of
the electron-phonon coupling with $V_{g}$ reduces $T_{c}$, while the
increase of $\epsilon _{F}$ (the charging appears according to experiment
only in the underdoped region) increases $T_{c}$. The overall effect is that
in the underdoped case the second tendency prevails, while in the overdoped
only the first exists. This explains the"dome" shape.

\section{Discussion and summary}

Pairing in one atomic layer thick two dimensional electron gas on a strongly
dielectric substrate by a single band of high energy longitudinal optical
phonons is considered in detail. The phonon band is assumed to be nearly
dispersionless with frequency $\Omega $. The polar dielectric $SrTiO_{3}$
exhibits such an energetic phonon mode and the 2DEG is created both when one
unit cell $FeSe$ layer is grown on its $\left( 100\right) $ surface and on
the interface with another dielectric like $LaAlO_{3}$. Both the adiabatic, $%
\epsilon _{F}>>\Omega $, and the nonadiabatic, $\epsilon _{F}<<\Omega $,
cases are considered and compare well with conventional weak coupling BCS
and with the local instantaneous interaction model (describing the
nonadiabatic regime close to the BEC crossover\cite%
{BEC2Drev,Melik,Randeria2D,Chubukov} still assuming that $\epsilon
_{F}>>E_{c}$, where $2E_{c}$ is the binding energy, so that the pairing is
the BCS type rather than BEC) respectively. The focus was however on the
intermediate region. The reason is that in several novel materials this is
precisely the case. In particular in high $T_{c}$ one unit cell $FeSe$ on $%
STO$ the Fermi energy is a bit smaller than the phonon frequency $\epsilon
_{F}=0.65\Omega $. In interface superconductors like $LaAlO_{3}/STO$
interfaces the ration is smaller $\epsilon _{F}/\Omega \sim 0.3$ still well
above the nonadiabatic limit. It turns out that in the crossover region the
critical temperature decreases very slowly as function of $\epsilon _{F}$,
up to $\epsilon _{F}=0.1\Omega $, see Figs 2 and 3, and only then drops fast
to zero.

The critical temperature was calculated within the weak coupling model of
superconductivity. The theory was applied to two different realizations of
such a system: 1UC$FeSe$/STO and $LaAlO_{3}/STO$ interfaces.

The numerical solution of the gap equation at $\alpha =0$ was compared with
an often utilized Kirzhnits dielectric approach for arbitrary ratio $%
\epsilon _{F}/\Omega .$ This comparison demonstrated excellent agreement
between two theories in non-adiabatic range while in adiabatic region the
Kirzhnits theory gives lower $T_{c}$ than the numerical solution of the gap
equation.

We conclude that, despite small electron concentration, very high critical
temperatures observed recently are consistent with the mostly phononic
mechanism already due to combination of two peculiar properties of the
system. First, since the optical phonon frequencies $\Omega $ are very large
and electrons reside in small pockets, $\Omega $ is larger than $\epsilon
_{F}$. Second, due to the huge dielectric constant of STO the Coulomb
repulsion is strongly suppressed inside the layer leading to small $\alpha $%
. The required value of the electron-phonon coupling in the superconducting
layer is $\lambda \sim 0.5$ in 1UC$FeSe$/STO and $\lambda \sim 0.2$ in
LAO/STO. In low $T_{c}$ LAO/STO the less suppressed Coulomb repulsion
results in significant reduction or even suppression of superconductivity. A
phenomenological model for dependence of $\epsilon _{F}$ and $\lambda $ on
electric field for the LAO/STO is proposed.

The main insight from this work therefore is that small value of $\epsilon
_{F}$ is not an obstacle to achieve $T_{c}$ of order $0.1$ $\Omega $ as long
as $\lambda $ is sufficiently large and the Coulomb repulsion is effectively
suppressed by polarization of the 3D substrate.

\textit{Acknowledgements. }

We are grateful J.Wang, C. Luo, J.J. Lin, M. Lewkowicz, Y. Dagan for helpful
discussions. Work of D.L. and B.R. was supported by NSC of R.O.C. Grants No.
98-2112-M-009-014-MY3 and MOE ATU program. The work of D.L. also is
supported by National Natural Science Foundation of China (No. 11274018).

\appendix

\section{Derivation of the pairing equations}

We derive the Gorkov's equations within functional approach starting the
effective action Eq.(\ref{Ae}). The partition function as a functional of
sources $\chi _{p\omega }^{\sigma }$ is:

\begin{equation}
Z\left[ \chi \right] =\int_{\psi }\exp \left[ -A_{e}\left[ \psi \right]
+\int_{p\omega }\left( \psi _{p\omega }^{\sigma }\chi _{p\omega }^{\ast
\sigma }+\chi _{p\omega }^{\sigma }\psi _{p\omega }^{\ast \sigma }\right) %
\right] \text{.}  \label{A1}
\end{equation}%
The free energy, $\mathcal{F}\left[ \chi \right] =-\log Z\left[ \chi \right]
$, defines the effective action and the "classical fields" via%
\begin{eqnarray}
\mathcal{A}\left( \psi \right) &=&\mathcal{F}\left[ \chi \right]
+\int_{p\omega }\left( \psi _{p\omega }^{\sigma }\chi _{p\omega }^{\ast
\sigma }+\chi _{p\omega }^{\sigma }\psi _{p\omega }^{\ast \sigma }\right) ;
\label{A_2} \\
\psi _{p\omega }^{\sigma } &=&\frac{\delta \mathcal{F}\left[ \chi \right] }{%
\delta \chi _{p\omega }^{\ast \sigma }},\psi _{p\omega }^{\ast \sigma }=-%
\frac{\delta \mathcal{F}\left[ \chi \right] }{\delta \chi _{p\omega
}^{\sigma }}\text{,}  \notag
\end{eqnarray}%
where the sources are expressed via the first functional derivative of $%
\mathcal{A}$,%
\begin{equation}
\chi _{p\omega }^{\sigma }=-\frac{\delta \mathcal{A}\left[ \psi \right] }{%
\delta \psi _{p\omega }^{\ast \sigma }},\chi _{p\omega }^{\ast \sigma }=%
\frac{\delta \mathcal{A}\left[ \psi \right] }{\delta \psi _{p\omega
}^{\sigma }}\text{.}  \label{A3}
\end{equation}

The inverse propagators, the second derivatives, form a Nambu matrix:

\begin{eqnarray}
\Gamma _{p\omega q\nu }^{\sigma \rho } &=&\frac{\delta ^{2}\mathcal{A}}{%
\delta \psi _{q\nu }^{\rho }\delta \psi _{p\omega }^{\sigma }};\Gamma
_{p\omega q\nu }^{\sigma \rho }=\frac{\delta ^{2}\mathcal{A}}{\delta \psi
_{q\nu }^{\rho }\delta \psi _{p\omega }^{\sigma }};  \label{A4} \\
\Gamma _{p\omega q\nu }^{\sigma \ast \rho } &=&\frac{\delta ^{2}\mathcal{A}}{%
\delta \psi _{q\nu }^{\rho }\delta \psi _{p\omega }^{\sigma \ast }}\text{.}
\notag
\end{eqnarray}%
Green's functions also form a Nambu matrix,%
\begin{eqnarray}
G_{q\nu p\omega }^{\rho \sigma } &=&\left\langle \psi _{p\omega }^{\sigma
\ast }\psi _{q\nu }^{\rho }\right\rangle =-\frac{\delta ^{2}\mathcal{F}}{%
\delta \chi _{q\nu }^{\rho \ast }\delta \chi _{p\omega }^{\sigma \ast }};
\label{A5} \\
G_{q\nu p\omega }^{\rho \ast \sigma \ast } &=&-\frac{\delta ^{2}\mathcal{F}}{%
\delta \chi _{q\nu }^{\rho }\delta \chi _{\omega }^{\sigma }};  \notag \\
G_{q\nu p\omega }^{\rho \sigma \ast } &=&\left\langle \psi _{p\omega
}^{\sigma }\psi _{q\nu }^{\rho }\right\rangle =-\frac{\delta ^{2}\mathcal{F}%
}{\delta \chi _{q\nu }^{\rho \ast }\delta \chi _{p\omega }^{\sigma }}\text{.}
\notag
\end{eqnarray}%
The two Nambu matrices obey $\Gamma ^{AC}G^{CB}=\delta ^{AB}$, that
constitute the Gor'kov equations.

Let us now calculate $\Gamma $. The gaussian average first derivatives
assuming only anomalous averages, are

\begin{equation}
\chi _{p\omega }^{\sigma }=D_{p\omega }^{-1}\psi _{p\omega }^{\sigma
}-V_{p-p_{2},\omega -\omega _{2}}\psi _{p_{3}\omega _{3}}^{\kappa \ast
}\left\langle \psi _{p_{2}\omega _{2}}^{\sigma }\psi _{p-p_{2}+p_{3},\
\omega -\omega _{2}+\omega _{3}}^{\kappa }\right\rangle \text{.}  \label{A 6}
\end{equation}%
The second derivatives are:%
\begin{eqnarray}
\Gamma _{p\omega q\nu }^{\sigma \ast \rho } &=&\delta ^{\sigma \rho }\delta
^{\omega \nu }\delta _{pq}D_{p\omega }^{-1};  \label{A7} \\
\Gamma _{p\omega q\nu }^{\sigma \rho } &=&V_{q-p_{2},\omega -\omega
_{1}}\delta _{-p_{1}-p_{2}+q+p}\delta _{\omega -\omega _{1}-\omega _{2}+\nu
}\left\langle \psi _{p_{1}\omega _{1}}^{\sigma \ast }\psi _{p_{2}\omega
_{2}}^{\rho \ast }\right\rangle \text{.}  \notag
\end{eqnarray}%
Using the translation symmetry,
\begin{eqnarray}
\left\langle \psi _{p\omega }^{1}\psi _{q\nu }^{2}\right\rangle &=&\delta
_{\omega +\nu }^{{}}\delta _{p+q}F_{p\omega },  \label{A8} \\
\Gamma _{p\omega q\nu }^{\sigma \rho } &=&\left\langle \psi _{p\omega
}^{1\ast }\psi _{q\nu }^{2}\right\rangle =\delta ^{\sigma \rho }\delta
_{p+q}\delta _{\omega +\nu }^{{}}D_{p\omega }^{-1}\text{,}  \notag
\end{eqnarray}%
the equation $\Gamma ^{AC}G^{CB}=\delta ^{AB}$ becomes Eqs.(\ref{Gor1},\ref%
{Gor2}).

\section{Long range RPA screened Coulomb repulsion}

In equation Eq.(\ref{PRPA}) one detail was not presented: subtraction of the
neutralizing background. Since at nonzero frequency the screened repulsion
does not become short ranged, the neutralizing background should be taken
into account. For our purposed the jellium model suffices\cite{Fetter}. To
this end one need the infrared cutoff $L$. The results for sufficiently
large $L$ converge (numerical simulations were performed for $L=30\frac{%
\Lambda +\epsilon _{F}}{N_{\varepsilon }-1}$).

The electronic part of the kernel Eq.(\ref{PRPA}), in our units $\hbar
=m=\Omega $ (unit of length $\hbar /\sqrt{\Omega m}$),

\begin{eqnarray}
P_{\mathbf{p,k},\omega }^{RPA} &=&\ \frac{e^{2}}{\epsilon }\left\{ \frac{%
\epsilon \left\vert \mathbf{p-k}\right\vert }{2\pi e^{2}}+\frac{1}{\pi }%
\left( 1-\frac{\left\vert \omega \right\vert }{\sqrt{\omega
^{2}+v_{F}^{2}\left\vert \mathbf{p-k}\right\vert ^{2}}}\right) \right\} ^{-1}
\label{C1} \\
&&-\frac{2\pi e^{2}}{\epsilon L}\delta \left( \boldsymbol{p-k}\right) \text{,%
}  \notag
\end{eqnarray}%
transformed to polar coordinates (using the rotation invariance) and then
changing to the energy variable $\varepsilon _{p}=p^{2}/2-\epsilon _{F}$
results in%
\begin{eqnarray}
P_{\varepsilon _{1},\varepsilon _{2},n}^{RPA} &=&\frac{e^{2}}{\epsilon }%
\int_{\phi =0}^{2\pi }\frac{1}{A+2B}-\frac{2e^{2}}{\epsilon L}\delta \left(
\varepsilon _{1}-\varepsilon _{2}\right) ;  \label{c2} \\
A &=&\frac{\epsilon }{e^{2}}\left( \sqrt{2\left( \varepsilon
_{1}+\varepsilon _{2}+2\epsilon _{F}-2\sqrt{\left( \varepsilon _{1}+\epsilon
_{F}\right) \left( \varepsilon _{2}+\epsilon _{F}\right) }\cos \phi \right) }%
+\frac{\pi }{L}\right) ;  \notag \\
B &=&1-\frac{\left\vert \omega -\nu \right\vert }{\sqrt{\omega _{n}^{2}+4\mu
\left( \varepsilon _{1}+\varepsilon _{2}+2\epsilon _{F}-2\sqrt{\left(
\varepsilon _{1}+\epsilon _{F}\right) \left( \varepsilon _{2}+\epsilon
_{F}\right) }\cos \phi \right) }}\text{.}  \notag
\end{eqnarray}%

\end{document}